\newif\ifcheckpagelimits
\checkpagelimitsfalse

\ifcheckpagelimits
 \documentclass[nofootinbib,prl,aps,twocolumn,showpacs,showkeys,%
 amsmath,amssymb,superscriptaddress,reprint,floatfix]{revtex4-1}
\else
 \documentclass[prl,aps,twocolumn,showpacs,showkeys,%
 amsmath,amssymb,superscriptaddress,reprint,floatfix]{revtex4-1}
\fi

\usepackage{graphicx}
\usepackage{color}
\usepackage{units}

\usepackage{hyperref}
\hypersetup{
pdftitle={Phase Stabilization of a Frequency comb using trapped atoms}
anchorcolor=blue,
colorlinks=true,
citecolor=blue,
urlcolor=blue,
linkcolor=blue}

\newcommand{\ket}[1]{\ensuremath{\left|{#1}\right\rangle}}
\newcommand{\bra}[1]{\ensuremath{\left\langle{#1}\right|}}

\newcommand{\mumeter}{\ensuremath{\mu\mathrm{m}}}

\usepackage[usenames,dvipsnames]{xcolor} 

\newcommand\frep{\ensuremath{f_{\rm rep}}}
\newcommand\taurad{\ensuremath{\tau_{\rm rad}}}
\newcommand\taucoh{\ensuremath{\tau_{\rm coh}}}

\newcommand\fceo{\ensuremath{f_{\mathrm{ceo}}}}
\newcommand\fcw{\ensuremath{f_{\mathrm{cw}}}}
\newcommand\Yb{Yb$^+$}

\begin{document}

\title{Phase Stabilization of a Frequency Comb using Multipulse Quantum Interferometry}

\author{Andrea Cadarso}
\address{Instituto de F\'{\i}sica Fundamental, IFF-CSIC, Serrano 113b, Madrid 28006, Spain}
\address{Departamento de An\'{a}lisis Matem\'{a}tico, Universidad Complutense de Madrid, 24040 Madrid, Spain}
\author{Jordi Mur-Petit}
\address{Instituto de F\'{\i}sica Fundamental, IFF-CSIC, Serrano 113b, Madrid 28006, Spain}
\address{Instituto de Estructura de la Materia, IEM-CSIC, Serrano 123, Madrid 28006, Spain}
\author{Juan Jos\'e Garc\'{\i}a-Ripoll}
\address{Instituto de F\'{\i}sica Fundamental, IFF-CSIC, Serrano 113b, Madrid 28006, Spain}


\begin{abstract}
From the interaction between a frequency comb and an atomic qubit, we derive quantum protocols for the determination of the carrier-envelope offset phase, using the qubit coherence as a reference, and without the need of frequency doubling or an octave spanning comb.  Compared with a trivial interference protocol, the multipulse protocol results in a polynomial enhancement of the sensitivity ${\mathcal O}(N^{-2})$ with the number $N$ of laser pulses involved.  We specialize the protocols using optical or hyperfine qubits, $\Lambda$-schemes and Raman transitions, and introduce methods where the reference is another phase-stable cw-laser or frequency comb. 
\end{abstract}

\ifcheckpagelimits
\else
 \maketitle 
\fi

Quantum Physics has experienced a universally recognized~\cite{Nobel2012} progress in the control and observation of individual quantum systems. In this respect, trapped ions~\cite{Leibfried2003a,Haffner2008a} is one of the most mature setups, with unbeaten precision in the realization of single-~\cite{Brown2011a} and two-qubit~\cite{Benhelm2008a} unitaries and measurements~\cite{Myerson2008a,Burrell2010}, closely followed by neutral atoms~\cite{Olmschenk2010}. This spectacular progress underlies a number of ``spin-off'', such as the characterization of atomic properties using entanglement~\cite{Widera2004} or the development of quantum algorithms and protocols~\cite{Mur-Petit2011,Leibfried2012,ding_quantum_2012} for studying molecular ions. The synergy is even more advanced in the field of metrology, with accurate atomic clocks assisted by quantum gates~\cite{Schmidt2005a,chou_frequency_2010} or the use of atomic squeezing for enhanced magnetometry~\cite{wasilewski_quantum_2010,Napolitano2011a}.

Despite the exquisite precision of atomic, molecular and optical (AMO) systems, the control and detection timescales ($\sim \unit[10]{\mu{s}}$ to $\unit[10]{ms}$) prevented using these techniques for studying ultrafast processes. In this work, we show that the speed of AMO setups is sufficient to accurately stabilize the carrier-envelope offset phase (CEP) of a frequency comb (FC). CEP effects are relevant for few-cycle pulses, though effects in multicycle pulses have also been reported~\cite{Jha2011}. The first observation of CEP effects was reported in the spatial asymmetry of above-threshold ionization from Kr gas~\cite{Paulus2001} and in x-ray emission from Ne~\cite{Baltuska2003}. The direction of photocurrents injected in semiconductors is also controlled by the CEP phase~\cite{Fortier2004,Roos2005} and the absolute CEP of single pulses was recently measured~\cite{Wittmann2009}.  The study of the CEP has been generally centered on its spectral components~\cite{Walmsley2009}, while only a few reports have addressed time-domain measurements of the relative phase of successive pulses in a train~\cite{Xu1996,Osvay2007}. The methods presented below follow this less-beaten path.

Let us introduce the notion of ``multipulse quantum interferometry'' (MPQI), where an atom acts as a nonlinear, fast-response detector that efficiently measures the differences between ultrashort laser pulses. Modelling the atom-pulse interaction as a sequence of unitaries, $\{U_i\}_{i=1}^N$, through a suitable reordering of the pulses, additional gates and measurements, we build protocols that accurately determine the differences among the pulses, or the properties of individual pulses themselves. Compared with cw laser interferometry, this approach provides a polynomial enhancement of the sensitivity because a single atom accumulates many interferometric events.

\begin{figure}[t]
  \centering
  \includegraphics[width=\linewidth]{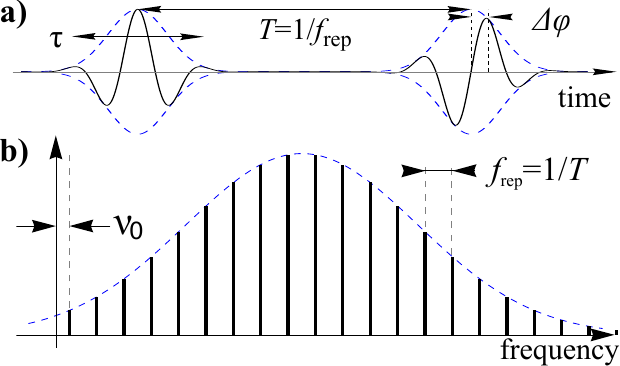}
  \caption{(a) Electric field amplitude (solid line) and envelope (dashed) of a pulsed laser with period $T$ and pulse-to-pulse phase difference $\Delta{\phi}$. %
(b) Associated spectra: a broad peak for one pulse (dashed) and a modulated comb for a pulse train with repetition rate $f_{\text{rep}}$ (solid). The frequency offset $\nu_0$ depends on the pulse-to-pulse phase difference $\Delta{\phi}$.}
  \label{fig:comb}
\end{figure}

A direct application of MPQI is the characterization and stabilization of a frequency comb~\cite{Udem2002,ye_femtosecond_2005}. This device produces a train of laser pulses with a fixed duration, $\tau$, and a regular spacing, $T$ [cf.~Fig.~\ref{fig:comb}a]. Stabilizing a comb is ensuring that the offset frequency, $\nu_0$, remains a constant and well-known value, and that the spectrum is a collection of regularly spaced teeth with frequencies $f_n = n/T+\nu_0$ [Fig.~\ref{fig:comb}b]. Haensch and Hall solved this problem~\cite{Reichert1999,Jones2000} in frequency space, interferometrically comparing different teeth in the limit of many pulses. Note that this requires a comb whose spectrum spans at least an octave, or broadening the light with a nonlinear fiber. This stabilization enables direct frequency comb spectroscopy, accurately revealing the atomic level structure of neutral atoms \cite{marian_united_2004, witte_deep-ultraviolet_2005} and ions \cite{wolf_direct_2009,wolf_direct_2011}.

We rather work on the time-domain image of the pulse train. The effect of the offset frequency is to change the CEP from pulse to pulse, $\phi_{n+1} - \phi_n = \Delta \phi = \nu_0 T$ [cf.~Fig.~\ref{fig:comb}a]. To address the problem of comb stabilization we will use MPQI, designing protocols that detect the phase difference between pulses with the greatest accuracy possible. We start by proposing a simple two-level protocol for consecutive pulses in a low intensity regime (1A) and in a $\theta \simeq \pi$ regime (1B). We further this study by introducing analogous protocols for delayed sequences of pulses which display an enhanced sensitivity (protocols 2A and 2B). Afterwards, in order to minimize spontaneous emission, we  describe equivalent protocols using Raman schemes. Finally, we present a discussion of experimental errors and the achievable sensitivities in practical implementations. The resulting methods do not require an octave-spanning comb, broadening or frequency doubling. They are thus useful for a wider variety of lasers, demand less power, and may profit from the ever-growing precision in atomic interferometry.

\paragraph{Single-pulse unitary.-}
We start by determining the unitaries associated to each laser pulse and how they depend on the CEP, $\phi_n$. The interaction of multilevel atoms with a frequency comb was studied previously \cite{felinto_theory_2009}. We model this interaction in the Rotating Wave Approximation (RWA) in order to produce analytical results\ \footnote{We can develop the MPQI protocols using the numerical expressions of the unitary operators when the RWA does not apply}
\ifcheckpagelimits\else
\begin{equation}
  H_{\mathrm{RWA}} = \tfrac{1}{2}(\omega_{at}-\bar{\omega})\sigma_z + s(t)\left(e^{-i\phi_m}\sigma^+ + \mathrm{H.c.}\right),
  \label{eq:RWA}
\end{equation}
\fi
Here, $m$ is the pulse index, $s(t)\geq 0$ is the pulse envelope, $\bar{\omega}=2\pi\bar{\nu}$ is the comb carrier frequency, $\omega_{at}$ is the atomic transition frequency ($\hbar=1$ throughout), there is an unknown phase $\phi_m$ for each pulse, and $\sigma_{x,y,z}$ are the Pauli matrices. The RWA works for pulses which contain $\geq 30$ periods of the carrier frequency, $\tau \geq 30/\bar{\nu}$~\cite{EPAPS}, and allows us to explicitly write the pulse unitaries
\ifcheckpagelimits\else
\begin{equation}
  U_m = \cos\left(\frac{\theta_m}{2}\right)+ i \sin\left(\frac{\theta_m}{2}\right) \sigma_{\phi_m}
  = e^{-i\phi_m\sigma_z} U_0e^{i\phi_m\sigma_z},
\end{equation}
\fi
in terms of the total Rabi flip angle of a single pulse, $\theta_m = 2 \int_{-\tau/2}^{\tau/2} s(t) dt$, with $\sigma_{\phi_m}  = \cos(\phi_m)\sigma_x + \sin(\phi_m)\sigma_y$. In what follows, we assume that the comb is almost resonant, $\bar{\omega}\simeq\omega_{at}$, and has uniform intensity, i.e. $\theta_m=\theta$. These assumptions imply that we only need to stabilize the pulse-to-pulse phase difference $\Delta\phi$.

\begin{figure}[t]
  \centering
  \includegraphics[width=\linewidth]{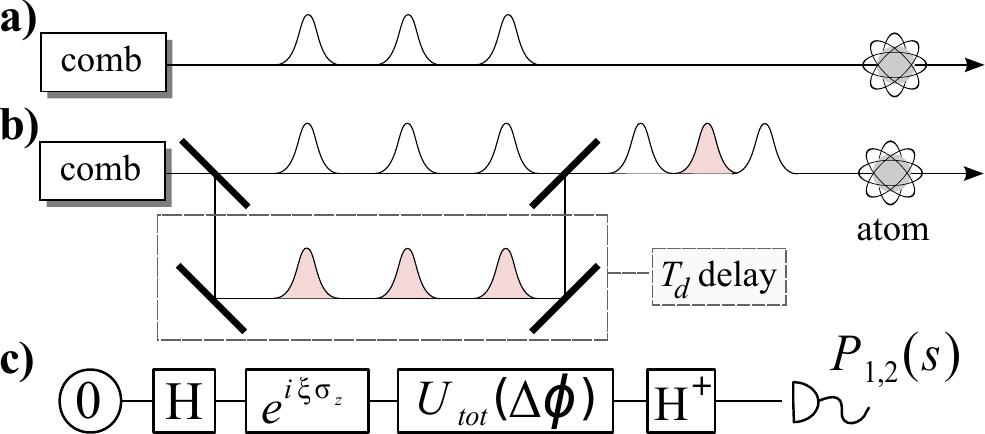}
  \caption{Comb phase measurement setups. A trapped atom interacts with (a) one train of pulses or (b) two trains with a delay $T_d$. (c) Additional gates and a final state interrogation build up a generalized Ramsey interferometry protocol to estimate $\Delta \phi$ and $\theta.$}
  \label{fig:setup}
\end{figure}

\paragraph{Multipulse unitaries.-}
We want a protocol that efficiently detects the difference between a sequence of unequal pulses $U_{tot} = \prod_{i=1}^N U_i,$ and the ideal case $U_1^N$. Let us first assume an ideal qubit, seeking an ordering of pulses with which the fidelity $\vert \text{tr}(U_1^{N\dagger} U_{\text{tot}}) \vert$ decreases most rapidly with $N.$ The simplest protocol (1A) applies $N$ consecutive pulses [cf. Fig.~\ref{fig:setup}a] with low intensity, $\theta \ll 1$, on the qubit, which adiabatically follows the phase 
\ifcheckpagelimits\else
\begin{equation}
\label{eq:U1A}
U^{(1A)}_{tot} \approx \openone + i
 \theta \frac{\sin(N\Delta\phi)}{2\sin(\Delta\phi)} \left[e^{i(N+1)\Delta\phi} \sigma^+ + \mathrm{H.c.}\right]
 + \ldots
\end{equation}
\fi
Note how the pulse-to-pulse phase difference $\Delta \phi$ decreases the amplitude of the Rabi oscillations and can be measured. However, as we show later on, the functional dependence on $\Delta\phi$ implies a low sensitivity on the phase in practical implementations of the protocol.

We can do much better by changing the intensity regime to $\theta=\pi$ (protocol 1B), where each comb pulse can flip the state of the atom. Under these conditions, for an even set of pulses we get
\ifcheckpagelimits\else
\begin{equation}
  \label{eq:U1B-general}
  U_{tot}^{(1B)}  = \prod_i U_i = \exp\left[-2i\sum_{k=1}^{N/2}(\phi_{2k}-\phi_{2k-1}) \sigma^z \right] ,
\end{equation}
\fi
which for constant $\Delta\phi$ implies $U_{tot}^{(1B)} = \exp\left(-i N\Delta\phi \sigma^z \right)$. Now $\Delta\phi$ can be interferometrically detected with an enhancement proportional to the number of pulses, $N$.

It is obvious that the sensitivity (\ref{eq:U1B-general}) increases by maximizing the phase difference between consecutive pulses. To profit from this, we design a set of protocols that extract a sequence of $N/2$ pulses from the original pulse train, and delay them a time $T_{d} \gg T$. This sequence is then intercalated with the original one, cf.~Fig.~\ref{fig:setup}b, so that $\phi_{2k} = k \Delta \phi + \Delta\phi \: T_{d} / T$ and $\phi_{ 2k-1} = k \Delta \phi$. Introducing this sequence in Eqs.~(\ref{eq:U1A}, \ref{eq:U1B-general}) we obtain respectively the unitaries corresponding to protocols 2A (for $\theta \ll 1$) and 2B ($\theta \sim \pi$). In particular, the unitary corresponding to protocol 2B is
\ifcheckpagelimits\else
\begin{equation}
  \label{eq:U2B}
  U_{tot}^{(2B)} = \exp\left(-i \sigma^z \Delta\phi  \times N {T_{d}}/{T} \right) \:,
\end{equation}
\fi
with an additional enhancement factor, $N_d=T_{d}/T$. This is optimal with respect to any rearrangement of the pulses, using each pulse only once.

\paragraph{Interferometry and sensitivity.-}
We now transfer the information of the acquired phase to the measurable populations of the atomic states. For this, we complete the previous unitaries with additional operations and measurements that enable estimating $\Delta\phi$ and $\theta$. Out of $2M$ atoms, $M$ are subject to the following steps [cf.~Fig.~\ref{fig:setup}c]: (i) initialization to the ground state, $\ket{0}$, (ii) apply a $\pi/2$ rotation (which could be either $\exp(i \sigma_x \pi/4)$ or a Hadamard gate) onto the ground state (iii) apply a reference phase $\xi$ onto the level $\ket{1}$, (iv) let the atom interact with the comb as described before, (v) undo the $\pi/2$ rotation of step (i) and measure the state of the atom, $s\in\{0,1\}$. The measurement outcome is described by the probability distribution, $P_1(s|\theta,\Delta\phi)$. For the remaining $M$ atoms we skip (ii), obtaining the distribution $P_2(s|\theta,\Delta\phi)$. We remark that we need no phase coherence between the comb and the lasers that implement the $\pi/2$ rotations. The reference phase, $\xi$, is computed a priori to maximize the sensitivity of $P_{1,2}$ to the $\Delta\phi$.

The functions $P_1$ and $P_2$ convey all the information accessible in the lab: from the measurements of $s$ in $P_1$ and $P_2$ experiments, one should compute different estimators and use them to infer the values of $\theta$ and $\Delta\phi$, with uncertainties $\sigma_{\theta}$ and $\sigma_{\Delta\phi}$. Using error propagation and the Fisher information we obtain fundamental lower bounds and practical estimates~\cite{EPAPS} of the sensitivities ($\sigma_{\Delta \phi}^{-1}$ and $\sigma_{\theta}^{-1}$) of each protocol. As summarized in Table~\ref{table}, it is possible to build estimators of minimal variance for $\theta$ and $\Delta\phi$, which saturate the fundamental lower bounds. Moreover, we observe that all protocols but 1A improve over the standard statistical sensitivity, $\sqrt{M}$, thanks to the large number of pulses or to the use of pulses from well-separated times. In practice, both $N$ and $N_d$ span several orders of magnitude, providing a sensitivity comparable to the state of the art. 

\begin{table}[tb]
\begin{tabular}{|l|c|c|}
  \hline \textbf{Implementation} & $\theta\ll 1$ (A) & $\theta\simeq \pi$ (B) \\
  \hline
  \hline 2 levels, no delay (1) & 
  $\sqrt{M}$ &
  $ N \sqrt{M}$ \\
  \hline 2 levels, with delay (2) & 
  $ N_d \sqrt{M}$ &
  $N N_d \sqrt{M}$ \\
  \hline
\hline Raman, 1 delay (1) &
  $N_d \sqrt{M}$ &
  - \\
  \hline Raman, 2 delays (2) &
  $ \vert N_{d2} - N_{d1} \vert \sqrt{M}$ &
 $ N \vert N_{d2} - N_{d1} \vert  \sqrt{M}$ \\
  \hline
\end{tabular}
\caption{Sensitivities, $\sigma_{\Delta\phi,\theta}^{-1}$, of a set of $2M$ two- or three-level atoms to the protocols described in the text (1A, 1B, 2A, 2B). $N$ is the number of pulses in a sequence, which in the delayed cases are combined with $N$ pulses from a later time, $T_d=N_d T.$}
\label{table}
\end{table}

\begin{figure}[t]
  \centering
  \includegraphics[width=\linewidth]{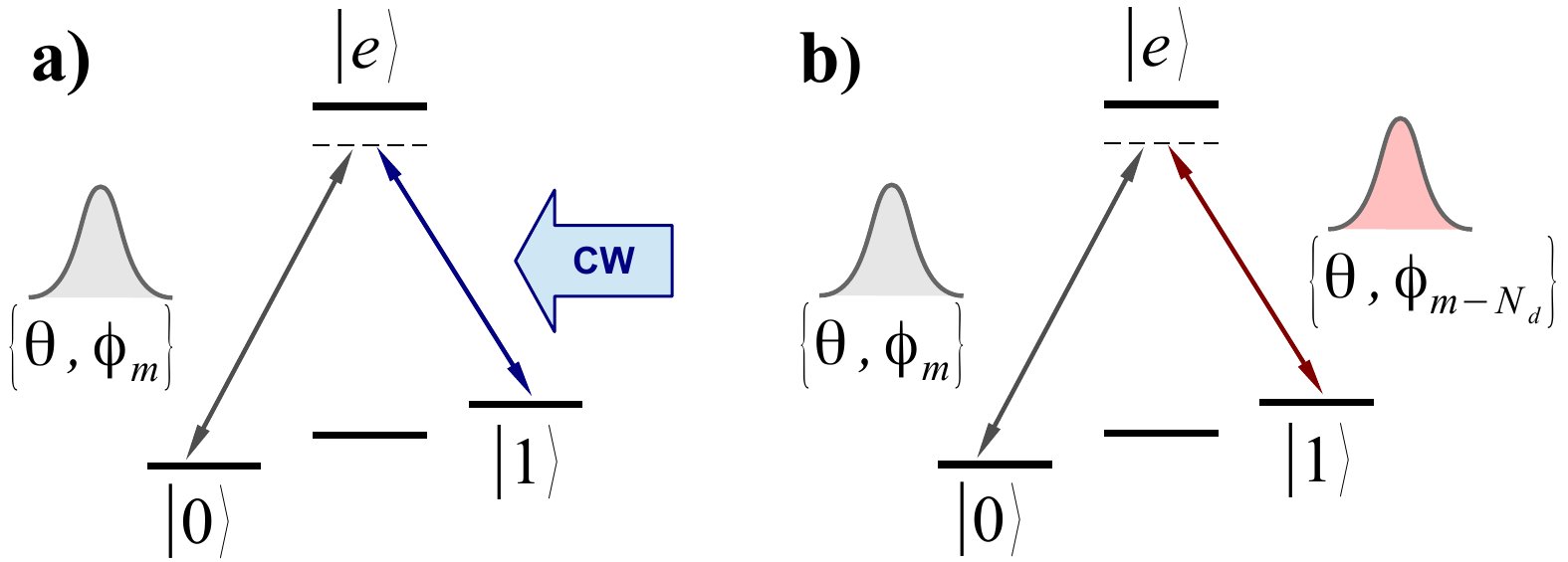}
  \caption{The $m$-th comb pulse interacts in a Raman setup with either (a) a cw laser signal or (b) another comb pulse. Controlling the polarization of the light and using the selection rules in atomic transitions we can ensure that each pulse or laser activates only one leg of the $\Lambda$ scheme.}
  \label{fig:raman}
\end{figure}

\paragraph{Three-level schemes.-}
In real atoms, if the qubit states $0$ and $1$ are dipole-coupled by a comb, spontaneous emission may severely limit the total interrogation time. One solution is to use dipole-forbidden
transitions 
restricted in practice to the $\theta \ll 1$ regime. An attractive alternative is the $\Lambda$-scheme in Fig.~\ref{fig:raman}, where two long-lived states, $\ket{0,1}$, talk via an intermediate level, $\ket{e}$. Applying combs or other lasers with orthogonal polarizations on the legs of the $\Lambda$-scheme, we can create effective Rabi oscillations between \ket{0} and \ket{1} while keeping a small population in $\ket{e}$ so that spontaneous emission is negligible.

A simple way to minimize spontaneous emission is to turn the $\Lambda$- into a Raman scheme, detuning the lasers that couple $\ket{0,1}$ with $\ket{e}$. Such Raman processes mix well with our algorithms. To start, if we have already stabilized the phase of a cw laser, we can combine it with the pulses from the comb [cf. Fig.~\ref{fig:raman}a]. This process enables an accurate determination of the CEP with respect to the cw source. The result is a sequence of effective unitaries with an average Rabi angle, $\theta'$, and a pulse phase $\phi'_m=\phi_m - \phi_{\rm ref}$, where $\phi_{\rm ref}$ is the phase of the stabilized source. The identifications $\theta \to \theta'$ and $\phi_m \to \phi'_m$ directly \textit{translate all protocols above to this new setup}. Likewise, one may combine the FC with a stabilized one [cf.~Fig.~\ref{fig:raman}b] and use our protocols to reconcile them.

A more interesting use of Raman transitions is to achieve self-referencing of the comb. For this, we use the scheme from Fig.~\ref{fig:raman}b, combining two pulses from the same comb, but with a relative delay, $T_d$, as in Fig.~\ref{fig:setup}b. This amounts to a self-referenced interferometric scheme based on time shifts, not requiring frequency shifting nor shearing \cite{Walmsley2009}. The phases of both pulses effectively combine in a nontrivial way in the unitary associated to the Raman process, $\phi'_m = \phi_m - \phi_{m - N_d} = N_d \Delta \phi$~\cite{EPAPS}. We can apply a sequence of $N$ pulse pairs with an effective angle $\theta'$ that should optimally lie around $N\theta' \simeq \pi/2$,
\ifcheckpagelimits\else
\begin{equation}
  \label{eq:1}
  U^{(1A,Raman)}= e^{-iN_d\Delta\phi\sigma_z} e^{i N{\theta' \over 2}\sigma_x} e^{iN_d\Delta\phi\sigma_z}
\end{equation}
\fi
and use Ramsey interferometry to measure both $\theta'$ and $\Delta\phi$. A generalization of protocols 2A and 2B is also possible using a linear optics circuit with two delay lines, so that each atom is hit by pairs of pulses with alternating phases $(\phi_m,\phi_{m-N_{d1}})$ and $(\phi_m,\phi_{m-N_{d2}})$. This leads to the sensitivities shown in the lower half of Table~\ref{table}.

Note that using Raman schemes demands the setup to be interferometrically stable up to a fraction of a wavelength. When a single pulse interacts with a two-level atom it does not matter whether the delay is a multiple of the comb period, or fails by a small amount, $\delta T = T_d - N_d T$ ($\vert \delta T \vert < T$). This is so because only the CEP enters the unitary and this only contains information on $\nu_0 N_d T.$ However, in Raman schemes, where two pulses overlap in time, their relative delay is a new parameter that influences the effective Rabi angle as well as the phase. In particular, the phase difference reads~\cite{EPAPS} $\Delta \phi' = \Delta \phi + \omega \delta T,$ with a contribution due to the interferometric path $c \delta T,$ which must be separately stabilized.

To remove the need for interferometric stability, we can use a different approach in which the comb only interacts with one transition, $\ket{1}\to\ket{e}$, performing $\pi$ rotations, while $\ket{0}$ is a dark state. The unperturbed and delayed pulses arrive closely in pairs, but without temporal overlap, implementing the sequence $\ket{1} \to -e^{i(\phi_m - \phi_{m-N_d})} \ket{1}$. Due to the lack of overlap, the delay errors drop and the effective operation is a phase gate in the qubit space. Spontaneous emission lowers the visibility and it is small because $\ket{e}$ is populated only a time $T_e = \mathcal{O}(\tau)$. Denoting by $\gamma$ the spontaneous decay rate of $\ket{e}$, we may afford $N=-\log(\epsilon)/\gamma T_e$ pulses before the visibility decreases by $\epsilon$. For a typical value $1/\gamma=8$~ns and a safe $T_e=100$~ps, visibility decreases just $10\%$ for $200$ pulses, sufficient to implement the last protocol in Table~\ref{table}.

\paragraph{Errors.-}
We can also account for ac Stark and Zeeman shifts in experiments.
In both cases, the effect can be modelled~\cite{EPAPS} as a random term in the Hamiltonian, $\epsilon(t)\sigma_z$, that makes the atomic levels fluctuate on timescales much longer than $\tau$.
This induces an uncertainty in $\Delta\phi$ of order $\sigma_\epsilon \times (t_{m+1}-t_{m})$, where $\sigma_{\epsilon}$ is the standard deviation of $\epsilon(t)$ from its (zero) average, and $t_m$ the arrival time of each pulse. This error is cancelled using spin-echo techniques~\cite{Kurnit1964} or, more directly, in protocols 2A and 2B, by calibrating the delays so that consecutive pulses arrive closely spaced but without overlap, say 10~ps apart. A pessimistic ac Stark shift $\sigma_\epsilon\sim 100\,$Hz then induces an error $\leq 10^{-9}$~rad in $\Delta\phi$.


Another source of error is temperature: when atoms move between pulses, they sample the laser's spatial variations of phase and intensity. We can eliminate such errors~\cite{EPAPS} (i) working in a Raman configuration which transfers no net momentum to the atom and (ii) ensuring the lasers are not tightly focused. These techniques allow working with sympathetically Doppler cooled ions in fast experiments ($\sim 1-10\,$ms from ion reset to detection).

The protocols discussed admit many implementations. For concreteness, we discuss here a setup with trapped ions, because of recent progress in connection with ultrafast lasers~\cite{Campbell2010a,mizrahi2013}. The long coherence times of ions, $t_{\rm coh} \sim1\,$s~\cite{Olmschenk2007a}, allow to consider trains of up to $t_{\rm coh}/T\sim 10^8$ pulses from a typical comb with $\frep\sim 100\,$MHz~\footnote{We discuss the maximum number of phase-coherent consecutive pulses in~\cite{EPAPS}.}. In the Raman schemes, with one ion and one delay line, this allows to detect CEP fluctuations $\delta \Delta\phi\sim 10^{-8}$~rad and calibrate the comb offset below $\delta\Delta\phi/T \sim 1\,$Hz, a remarkable precision for $1\,$s interrogation time! The numbers improve with a 2-delay Raman scheme, reaching $\delta\Delta\phi \sim 10^{-15}$ where error sources become relevant.
Precision decreases marginally, $\delta\Delta\phi \sim 10^{-5}-10^{-10}$, using faster duty cycles with $\sim 1\,$ms of interrogation time~\cite{EPAPS}.

\paragraph{Applications.-}
In practical applications, the phase differences will be large. To avoid it wrapping around $2\pi$, the number of pulses must be dynamically adjusted so that $N < 1/\Delta\phi$, increasing it only as the comb is better stabilized. Thus, measurement times cannot be longer than the typical time for the \textit{random} fluctuations in $\nu_0$. The precision limit is in practice set by the timescale at which we can provide useful feedback to the comb and not by the interferometric protocol.

We identify two frequency ranges where our protocol appears particularly useful. First, due to the technological and scientific interests of mid-infrared ($\lambda=2.5-25~\mumeter$) FCs~\cite{schliesser2012a}, we propose to use Ba$^+$ ions (that feature several narrow transitions around 2~\mumeter) to stabilize a visible or near-IR FC at $\Delta\phi=0$ so that difference-frequency generation from two of its teeth can produce a stabilized mid-IR FC. Secondly, Mg$^+$ presents various transitions around 280~nm which could be used to stabilize FCs in the near-UV, with application in high-harmonic generation and strong-field physics.  We discuss in~\cite{EPAPS} further details on current FC technologies, possible atom or ion stabilization systems, and a comparison between typical drift rates of an unlocked comb's frequency offset and the timescale of the atomic experiment.

Summing up, we presented several quantum inteferometric algorithms based on the idea that one atom may accumulate the effect of multiple laser pulses, computing their differences through the appropriate pulse ordering, intermediate gates and measurements. MPQI protocols provide a polynomial sensitivity enhancement with respect to conventional atom or Ramsey interferometry. MPQI can be used to detect temporal changes in the CEP of a FC because the unitary implemented by a single pulse is sensitive to both the intensity and the CEP, and not to the pulse arrival time. The schemes presented are particularly suitable for non-octave spanning combs with a low intrinsic phase noise, such as high-power Ti:Sapphire lasers where significant phase noise is introduced by amplification stages. Our protocols can be generalized beyond RWA and to characterize other properties of the comb, such as intensity fluctuations. We anticipate MPQI will enable new progress in fields as diverse as ultrafast science, frequency metrology and direct frequency-comb spectroscopy, or coherent control of molecular processes.

\ifcheckpagelimits
 \end{document}
\else
\begin{acknowledgments}
We acknowledge very long and fruitful discussions with Piet O. Schmidt.
This work has been funded by Spanish MINECO Project FIS2012-33022, CAM research consortium QUITEMAD (S2009-ESP-1594), COST Action IOTA (MP1001), a Marie Curie Intra-European Fellowship, and the JAE-Doc program (CSIC).
\end{acknowledgments}
\fi

\clearpage
\appendix

\section*{Supplementary material}

In the following pages we provide more details for some of the concepts which are spelled in the body of the paper:
\begin{itemize}
\item[\S{A}] Justification of the Rotating Wave approximation and its influence in the following calculations.
\item[\S{B}] Study of the Raman transitions and three level systems.
\item[\S{C}] Composition of unitaries for the different metrology protocols.
\item[\S{D}] Analysis of the sensitivity of the protocols using Fisher information and error propagation.
\item[\S{E}] Modelling of experimental errors and their influence in the quantum gates.
\item[\S{F}] Practical implementation with trapped ions and estimation of achievable sensitivies.
\item[\S{G}] Ultimate limits of these ideas based on pulse shaping techniques.
\end{itemize}

\tableofcontents

\section*{{\S}A. RWA and conventions}

\begin{figure}[t]
  \includegraphics[width=0.85\linewidth]{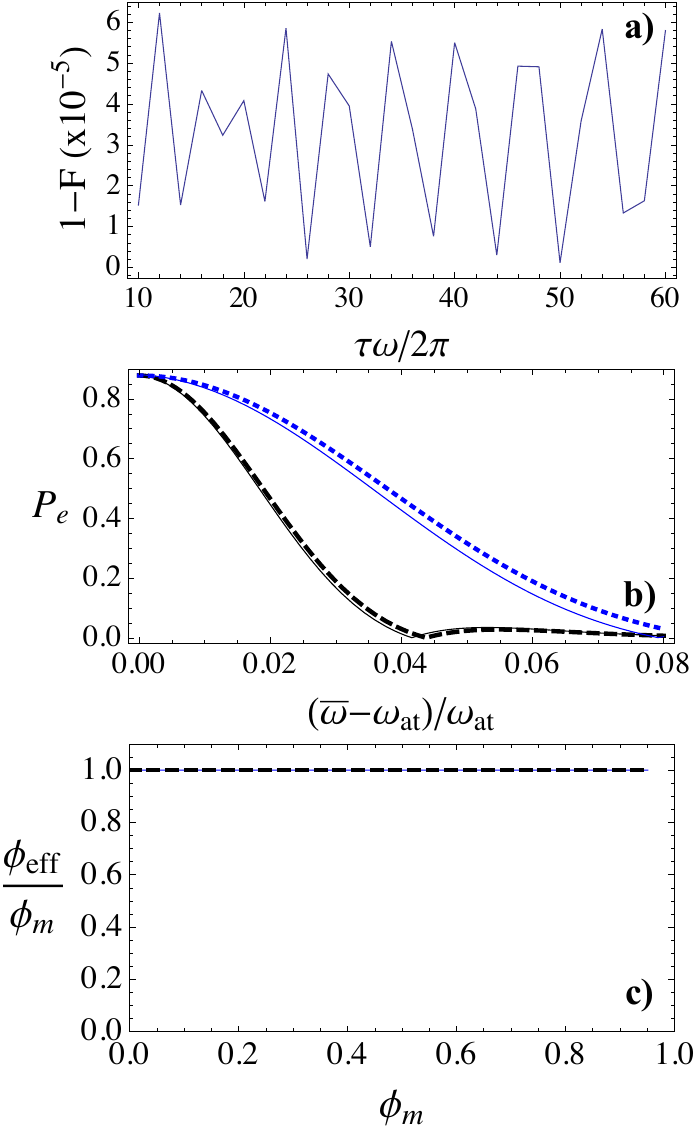}
  \caption{(a) Fidelity, $F$, between the unitary performed by the full model (\ref{eq:pulse-H}) and its description through the RWA~($H_{\mathrm{RWA}}$), as a function of the pulse duration, $\tau$, for a resonant pulse ${\bar\omega}=\omega_{at}$. (b) Qubit excitation probability for detuned pulses with $\tau=30\times 2\pi/{\bar\omega}$ (black) and $10\times 2\pi /{\bar\omega}$ (blue), and comparable Rabi frequencies. We show the exact solutions (solid) and the RWA (dashed). (c) Effective phase, $\phi_{\mathrm{eff}}$, of the unitary $U_m$ implemented by the full model (solid) and the RWA (dashed), as a function of the pulse phase, $\phi_m$, for resonant pulses ($\bar\omega = \omega_{at}$.}
  \label{fig:rwa}
\end{figure}

We model the atom interaction with a single laser pulse in the semiclassical limit
\begin{equation}
  \label{eq:pulse-H}
  H = \frac{\omega_{at}}{2}\sigma_z +
  s(t)\cos({\bar\omega}t + \phi_m) \sigma_x.
\end{equation}
Here, $s(t)\geq 0$ is the pulse envelope, and $\bar{\omega}=2\pi\bar{\nu}$ is the comb carrier frequency. The evolution under this Hamiltonian is described by a unitary operator that satisfies the Schr\"odinger equation $i\frac{d}{dt}W(t)=H(t)W(t)$. We are going to split out an evolution with the spin operator $\sigma_z$ as $W(t) = \exp(-i\bar{\omega} \sigma_z t/2)U(t)$. This operator evolves now according to
\begin{equation}
  i \frac{d}{dt}U = \frac{\omega_{at}-\bar{\omega}}{2}\sigma_z
  + s(t) (e^{i\bar{\omega}t+i\phi_m}+\mathrm{c.c.})
  (e^{i\bar{\omega}t}\sigma^+ +\mathrm{H.c.}).
\end{equation}
Assuming that the RWA is valid, in this Hamiltonian we can neglect the counter rotating terms, such as $e^{i2\bar{\omega}t+i\phi_m}\sigma^+$, and keep only those that are slowly varying. The result is $H_{\mathrm{RWA}}$ in the body of the paper.

But when is the RWA valid? We have performed numerical simulations of the evolution of the qubit under Eq.~(\ref{eq:pulse-H}), varying the duration of the pulse or number of oscillations it contains, as well as the intensity and detuning. In Fig.~\ref{fig:rwa}a we show the fidelity, $F$, of a resonant pulse, with a pulse area $\theta=\pi/2$, and a variable pulse length, $\tau$. The validity of RWA is also challenged by the inaccuracy of the control parameters, and in particular the driving frequency: as Fig.~\ref{fig:rwa}b shows, the unitary is affected by the detuning, and the differences between the RWA and the full model increase as the pulse length decreases. In practice this is not a problem, for we expect the detuning of the comb to be smaller than $1\%$. The main message is that for pulses above 30 oscillations, we are safe using the RWA Hamiltonian. 

As a final check, we show in Fig.~\ref{fig:rwa}c that the phase of the unitary is indeed proportional to the phase of the pulse. We obtain that $\phi_{\mathrm{eff}} = \phi_m + \phi_{\mathrm{AC}}$, where the carrier frequency appears in two different places: (i) determining the phase of the pulse, $\phi_m \sim \bar\omega t_m +\phi_0 \sim \nu_0 \times m \times T +\phi_0 \,\mathrm{[mod\,2\pi]}$, and (ii) in the AC Stark shift phase, $\phi_{\mathrm{AC}}$, that depends on the detuning $\omega_{\mathrm{at}}-\bar{\omega}$ and which becomes strictly zero for resonant pulses [Fig.\ \ref{fig:rwa}c]. In all other detuned cases, $\phi_{\mathrm{AC}}$ will either be the same for all pulses, in which case it will be eliminated by our algorithms, which are based on phase differences, or we will be able to take it into account with the error analysis from Sect.~\ref{sec:errors} below, by studying the fluctuations of the two-level frequency around the mean value $\omega_{at}$.

\section*{{\S}B. Raman transitions}

\begin{figure}[t]
  \centering
  \includegraphics[width=0.75\linewidth]{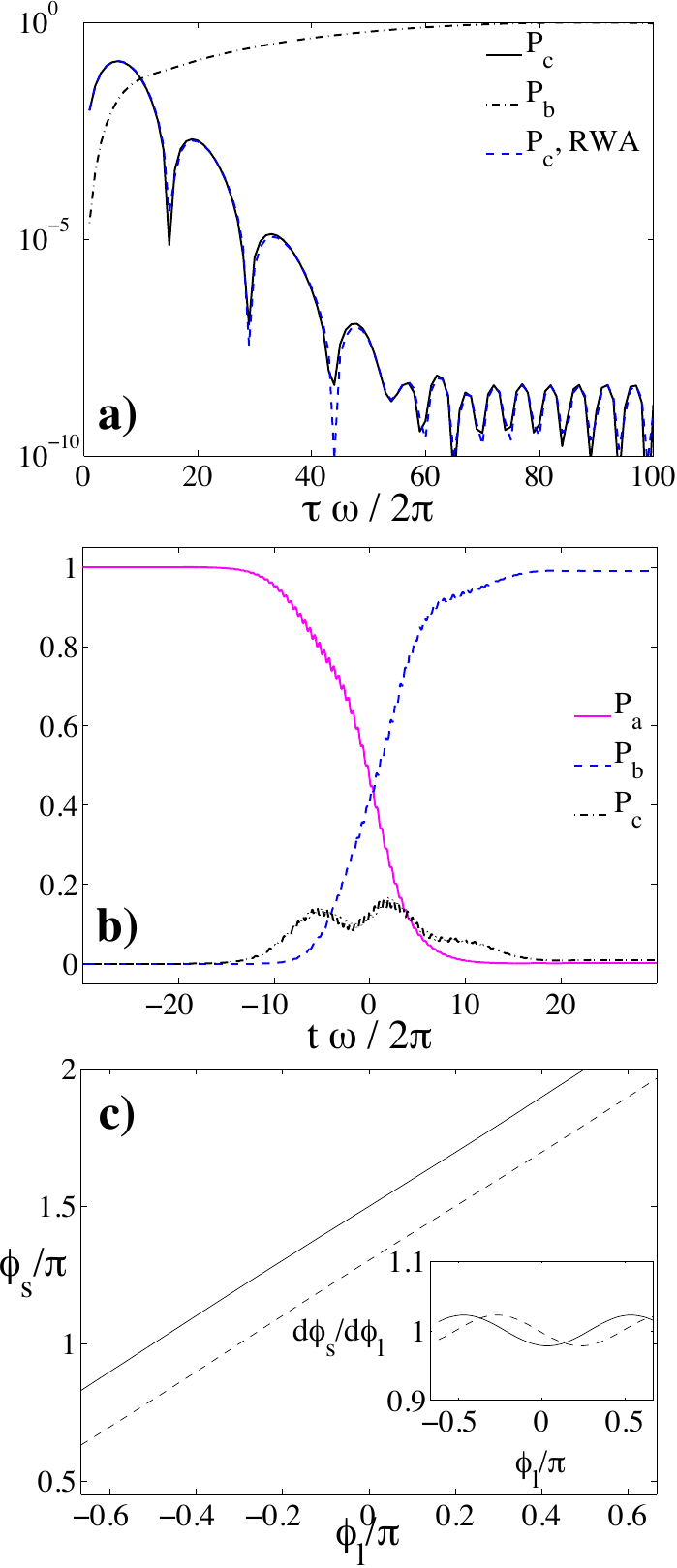}
  \caption{(a) Intermediate state population, $P_c$, and population of the state $b$, $P_b$, after a pulse of duration $\tau$, for $\omega=0.8\omega_{at}$ and Rabi frequency $\Omega=0.033\omega_{at}$. (b) Evolution of the state populations during a pulse with $\Omega=0.076\omega_{at}$, $\tau=40\pi/\omega$ and $\omega=0.8\omega_{at}$. (c) Relative phase between states $b$ and $a$ vs. the phase difference of the Raman pulse $\phi_l = \phi_2 - \phi_1$ in Eq.~(\ref{3level}) for $\Omega=0.016\omega,\, \omega=0.98\omega_{at}$ and $\tau=400\pi/\omega$. (Inset) Derivative $\frac{d\phi_s}{d\phi_l}$ confirming the monotonic behavior of $\phi_s$.}
 \label{fig:3level}
\end{figure}

Our Raman protocols are developed assuming that we can use ultrashort pulses to implement Raman transitions between two states, $a$ and $b$, mediated by a third one, $c$, which remains unpopulated at the end of the pulse. (Due to the very short duration of the interrogation sequence, it is not necessary that $c$ be unpopulated at all times, as is usual in STIRAP processes.) Importantly, we need that such operations implement the same quantum gates and carry the same phase information as the original designs. We are going to discuss both requirements and how they are achieved.

Note that Raman transitions with very short pulses have been demonstrated experimentally by the group of C. Monroe {\em et al.} in a series of works that implement quantum gates with trapped ions and pulsed lasers~\cite{Hayes2010,Madsen2006}. In these references, an interpretation based on Raman transitions induced by all the comb teeth is provided, but here we will discuss a different one.

For us the key aspect of a Raman transition is the fact that the intermediate state, $c$, is completely depopulated at the end of the process. In order for this to happen, we need that the energy of the final state is similar to the energy of the original one. Intuitively, this implies that the inverse of the duration of the process has to be smaller than $\delta = \omega_{at}-\omega$, the detuning of the laser from the atomic transitions $\{a,b\}\leftrightarrow c$, but larger than the difference $|\omega_{ac}-\omega_{bc}|$. As shown in Figs.~3a-b, this qualitative appreciation remains true even for rather extreme cases. In those exaggerated plots, we see that pulses with a detuning $\delta \sim 0.2 \omega_{at}$ work fine even when they only contain $10-20$  oscillations of the laser. In this regime the excited state $c$ is significantly populated during the pulse, but it has a population of less than $10^{-3}$ at the end.

The other aspect we demand from these pulses is the fact that they must carry information on the phase of the laser. To check this, we analyze the interaction between the three-level atom and the light using a simple Hamiltonian,
\begin{align}
  H &= s(t) \cos(\omega t+\phi_1)\ket{c}\bra{a} + \mathrm{H.c.} \label{3level}\\
  & + s(t) \cos(\omega t+\phi_2)\ket{c}\bra{b} + \mathrm{H.c} \nonumber\\
  &+ \omega_{at} \ket{c}\bra{c},\nonumber
\end{align}
which under the RWA becomes
\begin{align}
  H_{RWA}(\phi_1,\phi_2) &= s(t) e^{i\phi_1}\ket{c}\bra{a} + \mathrm{H.c.} \label{3levelrwa}\\
  & + s(t) e^{i\phi_2}\ket{c}\bra{b} + \mathrm{H.c} \nonumber\\
  &+ (\omega_{at}-\omega) \ket{c}\bra{c}. \nonumber
\end{align}
Note how $H_{RWA}(\phi_1,\phi_2)$ is related to $H_{RWA}(0,0)$ through a unitary transformation $\exp(-i\phi_l\sigma^z_{ab})$ in the $\{a,b\}$ subspace, with the relative phase $\phi_l = \phi_2-\phi_1$. In other words, according to the RWA the phase of the laser is mapped onto the relative phase between the states. The question is whether this behavior also follows from the original Eq.~(\ref{3level}). We have performed numerical simulations of the three level system in Eq.~(\ref{3level}) and the conclusions are: (i) There is always a small deviation between the real phase and the RWA approximation. (ii) This deviation decreases with decreasing detuning, as in the two-level system, an indication that it is due to the AC Stark shift effect. (iii) The actual phase experienced by the atom is a monotonic function of the laser phase, that is $\phi_s(\phi_l)$ grows with $\phi_l$. These properties are exemplified in Fig.~\ref{fig:3level}c for a case with $2\%$ detuning, where the deviations from the RWA are small, below $1\%$, but the nonlinear behavior is clear in the inset.

It would seem that, since we are striving for large accuracies in the stabilization protocol, errors of $1\%$ would be enough to discard the protocols. However, we have to remember that we are not actually measuring the absolute phase, but the phase difference between pulses. Hence, stabilizing $\phi_s$, which is a smooth, monotonic function of the laser phase, is equivalent to (and as accurate as) stabilizing $\phi_l$.

\section*{{\S}C. Analytical pulse estimates}

We summarize some of the arguments in the body of the article regarding the composition of pulses.

\subsubsection*{Pulse composition}

Let us start with the case of a sequence of pulses with $\theta\ll 1$ and a uniform carrier-envelope frequency mismatch, $\phi_m = m \Delta\phi$. The combination of pulses reads
\begin{align}
  U_{tot}^{(1A)}
  & = \prod_{m=1}^N e^{-i\phi_m\sigma_z} e^{i\theta/2 \sigma_x} e^{+i\phi_m\sigma_z}\\
  & = \cos\left(\frac{\theta}{2}\right) + i \sin\left(\frac{\theta}{2}\right) \sum_{m=1}^N e^{-i\phi_N\sigma_z}\sigma_x e^{i\Delta\phi\sigma_z} \nonumber \\
  & \simeq \openone + {i \frac{\theta}{2}} e^{-i(N+1)\Delta\phi} \frac{\sin(N\Delta\phi)}{\sin(\Delta\phi)}\sigma^+
  + \mathrm{H.c.} + {\mathcal O}(\theta^2)\nonumber
\end{align}
Note that in this context $\Delta\phi$ has the effect of a detuning and suppresses for long trains any excitation probability induced by the pulses. When we work around $\theta=\pi$ we obtain instead
\begin{align}
  U_{tot}^{(1B)} =  \prod_{m=1}^N i e^{-i\phi_m\sigma_z} \sigma_x e^{+i\phi_m\sigma_z} 
\end{align}
If we assume that the number of pulses is even, we can use the anticommutation $\sigma_x\sigma_z=-\sigma_z\sigma_x$ and
\begin{align}
  e^{-i\phi_m\sigma_z}  \sigma_x e^{+i\phi_m\sigma_z}  e^{-i\phi_{m-1}\sigma_z} \sigma_x  e^{+i\phi_{m-1}\sigma_z} \nonumber \\
  = e^{-2i(\phi_m - \phi_{m-1}) \sigma^z},  
\end{align}
recovering the formula
\begin{equation}
  U_{tot}^{(1B)}  = \exp\left[-2i\sum_{k=1}^{N/2}(\phi_{2k}-\phi_{2k-1}) \sigma^z \right]
\end{equation}
from the paper.

\subsubsection*{Some optimality considerations} 

We now prove that the sequence for protocol 2B (2 sequence of pulses split from the original train with a time delay) is optimal when our only resource is the comb laser.  As seen before, if we work around $\theta=\pi$ we obtain the analytical formula \begin{equation}
  U_{tot}^{(1B)}  = \exp\left[-2i\sum_{k=1}^{N/2}(\phi_{2k}-\phi_{2k-1}) \sigma^z\right]
\end{equation} and our protocol accumulates phase quite fast, about $\mathcal{O}(N N_d)$ for $N$ pulses, where $N_d$ depends on the delay. It is possible to prove that for any rearrangement of the same set of pulses (that is, with the same phases and intensity as before) this is the largest accumulation that can be detected.

If $\sigma$ is a permutation for a certain arrangement of initial pulses, we can use the analytical expression for the arbitrary product of a train of pulses with different phases to compute product of the unitaries after the permutation
\begin{equation}
\prod_{i=1}^{M}U_{\sigma(i)} = e^{2i\sum_{i=1}^{M}(-1)^{\sigma(i)}\phi_{\sigma(i)}\sigma^z}\sigma_{x}^{M}
\end{equation}

It is possible to find all permutations $\sigma$ such that they maximize $\vert \sum_{i=1}^{M}(-1)^{\sigma(i)}\phi_{\sigma(i)} \vert.$ Suppose the original pulses are ordered in terms of their carrier-envelope phase $\phi_n$, then it is quite straightforward to see how to construct optimal rearrangements of these pulses. Consider $\alpha$ a permutation of the first half of pulses and $\beta$ a permutation of the second half of pulses. Then, the optimal set of rearrangements will be those formed by pulses labelled according to their carrier-envelope phase as $\{ \phi_{\alpha(i)},\phi_{\beta(N/2 + i)}\}_{i \in \{1, \ldots, N/2\}}$ or of the form $\{\phi_{\beta(N/2 + i)}, \phi_{\alpha(i)}\}_{i \in \{1, \ldots, N/2\}}.$

In particular, our proposed protocol corresponds to $\alpha$ and $\beta$ being the identity permutation. This protocol accumulates the largest possible amount of phase after the action of the pulses onto the ion.

\subsubsection*{The fastest phase-accumulation protocol: phase referencing}

If we allow for more gates, performing unitaries in between the pulses, we can measure not only the phase difference, but also the total sum of the carrier-envelope phases $\sum_i \phi_i.$ In order to do so, the new set of gates and unitaries, considered in order, would be $\{\sigma_x, U(\phi_1), \ldots, \sigma_x, U(\phi_N) \},$ for which the overall product is
\begin{eqnarray*}
&& \prod_{i=1}^{M} \sigma_x U(\phi_{\sigma(i)}) = \prod_{i=1}^{M} \sigma_x e^{-i\phi_i \sigma_z } \sigma_x e^{ i\phi_i \sigma_z } \\ &=& \prod_{i=1}^M e^{i\phi_i \sigma_z } \sigma_x^2 e^{ i\phi_i \sigma_z } = \prod_{i=1}^M e^{i\phi_i \sigma_z } e^{ i\phi_i \sigma_z } \\ &=&  \prod_{i=1}^M e^{2i\phi_i \sigma_z } = e^{2i \sum_i \phi_i \sigma_z} 
\end{eqnarray*}
where we use both $\sigma_{x}e^{-k\sigma_{z}}=e^{k\sigma_{z}}\sigma_{x}$ and $\sigma_x^2 = Id$.

Note however this protocol demands $\sigma_x$ gates in between the pulses. Since the phase of these gates is stable, we can thus view this extra protocol as the referencing of the comb to the device that implements the $\sigma_x$ gates, which can itself be a laser or a microwave beam, in the case of hyperfine qubits.

\section*{{\S}D. Fisher information and sensitivity}

We are interested in estimating the sensitivity of the interferometric protocols that we have developed with respect to changes in the parameters they depend on. A measure of the information that one can extract about one or several parameters from a given probability distribution is the so-called Fisher Information \cite{Helstrom1976, Holevo2011}. 

In our protocols, we want to estimate the intensity of each of the pulses $\theta$ (considered constant throughout the whole experiment) and the pulse-to-pulse phase difference $\Delta \phi$. They will be related to some physical observables which measure the population of the excited state after applying certain protocols. The precision of the parameters $\theta$ and $\Delta \phi$ is determined by the fluctuations of these observables and their variance can be obtained using standard error propagation theory. The Fisher Information will yield a measure of the available precision in the estimation of the parameters. Also, the variance of the estimation of a given parameter will be limited by the Cramer-Rao bound \cite{Cramer1999}, which sets the ultimate limit for the precision that we can achieve. 

Let us see how to compute both the Fisher Information and the Cramer-Rao bound. Let ${\bf X}$ be a sample of observations with joint probability distribution given by $P({\bf X}\vert \mathbf{k})$ depending on a vector parameter $\mathbf{k } = \left( k_1, k_2, \ldots, k_i \right)^T$ and $h(\mathbf{k}),$ a real valued function of $\mathbf{k}$. Then, under suitable regularity conditions (see \cite{Helstrom1976, Holevo2011}), for any unbiased estimator $\hat{h}(\mathbf{X})$ of $h(\mathbf{k})$ $$\text{Var}(\hat{h}) \geq \mathbf{\delta}^T [I(\mathbf{k})]^{-1}\mathbf{\delta}$$ where $\mathbf{\delta}$ is the vector of derivatives of $h(\mathbf{k})$, i.e.
\begin{equation}
\mathbf{\delta} = \left( \frac{\partial}{\partial k_1} h(\mathbf{k}), \frac{\partial}{\partial k_2} h(\mathbf{k}), \ldots, \frac{\partial}{\partial k_i} h(\mathbf{k}) \right)
\end{equation}
and the matrix $I(\mathbf{k})$ is the Fisher Information matrix with $(i,j)$th element 
\begin{align}
	I_{ij}(\mathbf{k}) &= \mathbf{E} (S_i(\mathbf{X}) S_j(\mathbf{X})) \\
	&= - \mathbf{E} \left(\frac{\partial^2}{\partial k_i \partial k_j} \log P(\mathbf{X} \vert \mathbf{k})\right)
        \nonumber
\end{align}
where  $\mathbf{E}(\cdot)$ denotes the expectation value and
\begin{equation}
S_i(\mathbf{X}) = \frac{\partial}{\partial k_i} \log P(\mathbf{X} \vert \mathbf{k}).
\end{equation}

In particular, if $\hat{k}(\mathbf{X})$ is an unbiased estimator of the $r$th parameter $k_r,$ then, under the same regularity conditions $$\text{Var}(\hat{k}_r) \geq J_{rr}(\mathbf{k}),$$ where $J_{rr}(\mathbf{k})$ is the $r$th diagonal element of the inverse matrix $[I(\mathbf{k})]^{-1}.$ 

We have both used standard error propagation theory and computed the Fisher Information as explained before in order to get fundamental lower bounds and practical estimates of each of the proposed protocols which, for $\Delta \phi$ and $\theta$, we have shown in Table~I of the main text.

\section*{{\S}E. Experimental errors}

\subsubsection*{Dephasing}

Continuing the discussion from Sect.\ref{sec:rwa}, we want to clarify with greater detail our estimates of the errors induced by small detunings and energy shifts. From the theoretical point of view we consider a general situation in which we split the energy levels of the atom into two contributions: the average spacing, $\omega_{at}$, and random fluctuation of zero mean, $\epsilon(t)$, on top of it:
\begin{equation}
  H = \frac{\omega_{at}+\epsilon(t)}{2}\sigma_z +
  s(t)\cos({\bar\omega}t + \phi_m) \sigma_x.
\end{equation}
We assume that $\epsilon(t)$ may be random but always smoothly varying, so that it will not only remain constant within a single pulse, but it will also change slowly between consecutive pulses, $\frac{d}{dt}\epsilon(t) \ll t_{m+1}-t_m$.

Evolution then splits into two consecutive operations. Before the arrival of the pulse, $s \simeq 0$, the atom evolves freely with fluctuating energy levels, while during the pulse $\epsilon(t) \simeq \epsilon_n$ and the external field is approximately constant. In other words, the evolution after the waiting period and the pulse may be written as a product of two unitaries, $U_n = U_{\mathrm{pulse},n} U_{\mathrm{free},n}$. The free evolution is not significantly affected by the random fluctuations
\begin{equation*}
  U_{\mathrm{free},n} = \exp \left ( i \int_{t_{m-1}+\tau/2}^{t_m-\tau/2} [\omega_{at} + \epsilon(t)] dt\right) \simeq
  \exp(-i \omega_{at} T).
\end{equation*}
During the laser pulse, however, the interaction between the atom and the light is ruled by the equations from Sects.\ref{sec:rwa} and \ref{sec:raman}. In these sections we have seen that the effect of any detuning is (i) an extremely small change of the excitation probability, $\theta_n$, and (ii) an equally small AC Stark shift that changes the effective phase seen by the atom.

If these deviations in the rotation angles remain constant within consecutive pulses, they will be taken into account and suppressed by our algorithms. In other cases, they will contribute to the errors in the estimation of the CEO. Assuming that we can bring the pulses close together, so that the time lapse between pulse arrivals is comparable or smaller than the timescale of the fluctuations, we will find that the difference between two consecutive Stark shifts is proportional to the difference in arrival times and to the standard deviation of such external field fluctuations, $\sigma_\epsilon \times (t_{m+1} - t_m)$.

\subsubsection*{Temperature}

Temperature can also induce dephasing: if the atom is not still enough and it has time to move between pulses, it will see different phases of the pulse which depend on the distance traveled as $2\pi \Delta x/\lambda$. There are various ways to address this issue. We can make a simple estimate for a free atom in space, assuming that it is Doppler cooled. The temperature of the atom is
\begin{equation}
  k_B T \simeq \hbar \Gamma,
\end{equation}
where $\Gamma$ is the natural linewidth of the cooling transition. Let us pessimistically assume that all this energy goes to the kinetic part, $\frac{1}{2}mv^2$, giving us an average velocity
\begin{equation}
  v \simeq \sqrt{\frac{2k_B T}{m}} \simeq \sqrt{\frac{2\hbar \Gamma}{m}}.
\end{equation}
From this we can estimate the phase errors as
\begin{equation}
  \delta\phi \simeq \frac{2\pi}{\lambda} v (t_{m}-t_{m-1}).
\end{equation}
Also pessimistically we will take $\Gamma \simeq 200$ MHz, and a light atom such as Be, obtaining $v\sim 5$ m/s, which for a pulse separation of 10~ps gives $10^{-3}$ (or actually a bit larger if we consider the additional velocity due to the photon recoil).

The situation is very much improved when we arrange the lasers to work in a co-propagating Raman configuration such that there is no net momentum transfer to the atom. This is indeed a solution to the previous problem for, on each pulse, the spatially dependent phase which is acquired from one laser, $\vec{k}\vec{x}$, is cancelled by that of the other laser, $-\vec{k}\vec{x}$, making the whole process independent on the position of the atom. In this favorable circumstance, the only effect that temperature may have is when the intensity of the laser varies spatially. However, by using a laser beam which is not too tightly focused and confining the atoms to a small region, the effect of this inhomogeneity may be safely neglected.

\section*{{\S}F. Experimental sensitivities with trapped ions}

As mentioned earlier, trapped atomic ions~\cite{Leibfried2003} constitute one of the most mature systems in the implementation of Quantum Information Processing and Communication (QIPC) protocols and technologies. Precision records have been achieved in the realization of single-qubit~\cite{Brown2011} and two-qubit~\cite{Benhelm2008} unitaries and measurements~\cite{Myerson2008,Burrell2009}, even reaching the threshold for fault-tolerant quantum error correction protocols~\cite{Chiaverini2004}.
In the last couple of years, fantastic progress in the controlled interaction between trapped ions and ultrafast lasers has been achieved by the group of C.~Monroe at U.~Maryland, with the demonstration of two qubit entanglement in the weak-field ($\theta\ll1$), many-pulses regime~\cite{Hayes2010}, and also in the strong-field ($\theta \sim \pi$), few-pulses regime~\cite{Senko2012}, where logic gates faster than the trap oscillation period become accessible~\cite{Campbell2010,Garcia-Ripoll2003}.
Because of this, we think that the technology required to implement the phase-stabilization protocol that we propose is already available.

\begin{table}[tb]
\begin{center}
\begin{tabular}{|c|c|c|c|}
  \hline \frep (MHz) & $T$ (ns) & $\tau$ (ps)  & $T_d$ (ps)
  \\
  \hline
	100 & 10 & 10 & 10-100
  \\
  \hline
\end{tabular}
\end{center}
\caption{Typical parameters of a frequency comb~\cite{Campbell2010,Senko2012}.}
\label{tab:comb}
\end{table}

Typical parameters for the frequency comb are listed in Table~\ref{tab:comb}. Pulses with a duration $<1$~ps are nowadays easily accessible. On the other hand, one has to keep in mind that a pulse duration $\tau$ effectively limits the possible pulse delay times to $T_d > \tau$ in order to avoid an overlap of the electric fields corresponding to different pulses: depending on their polarizations, this may lead to several unwanted effects, from excitation of motional sidebands to total cancellation of the Raman transition~\cite{Campbell2010}; in either case, the action of a pulse pair on the qubit would still be described by a unitary transformation $U_i$, but not the ones we have written down earlier, so that our model would break down. Therefore, we will stick to a comb with pulses of 1-$10\,$ps.

In the following, we present details of our calculations to estimate the achievable sensitivity enhancements for the protocols introduced in the main text of the article. At the end, we present an estimation of the ultimate sensitivity limit that can be reached with pulse shaping techniques.

\subsubsection{Sensitivity enhancements with two-level protocols}

Protocols 1A to 2B consider direct transitions induced only by the comb laser that we want to study. In practice, there are two ways that this can be achieved: dipole and quadrupole transitions. Dipole transitions are, for instance, the $^2S_{1/2}$--$^2P_{1/2,3/2}$ lines in Yb${}^+$~\cite{Madsen2006}. These transitions have a typical linewidth of a few tens of MHz, which is comparable to \frep. This implies that the time between consecutive pulses could be shorter than the lifetime of the excited state of the ion. A possible solution to this problem will be presented later on in Sect.~\ref{sec:ultimate}.

An alternative is to rely on quadrupole transitions, as provided by the Ca$^+$ ion using as qubit states the electronic states $|S_{1/2}\rangle$ and $|D_{5/2}\rangle$~\cite{Haffner2008}. The excited level now has a radiative lifetime $\taurad \sim 1\,$s which is favorable to implement our ideas. The downside of quadrupole transitions is their lower coupling strength, which demands a more powerful laser to excite them. In practice, depending on the laser, this might imply that we have to work in the limit $\theta\ll 1$ (protocols 1A and 2A) but we will forget this in following discussion.

Following Table~\ref{tab:comb}, let us consider a frequency comb composed of 10-ps pulses with $\frep=100$~MHz.  This pulse duration is much shorter than the trap oscillation period ($1~\mu$s for a typical $\omega_{\rm trap}=2\pi\times 1$~MHz rf Paul trap) and allows us to disregard the motional state of the ion in the trap [cf.~Eq.~\eqref{eq:pulse-H}] as well as its micromotion, which may affect the performance of coherent protocols at longer times~\cite{Madsen2006}.  Such a frequency comb, and a typical ion coherence time for the electronic qubit states of $^{40}$Ca$^+$, $\taucoh \gtrsim 10$~ms~\cite{Haffner2008}, allow $\taucoh\frep \gtrsim 10^6$ pulses to go through the ion before decoherence becomes relevant. We take a conservative estimate of $T_{\mathrm{inter}}=1\,$ms for the time during which the ion is accumulating information on $\Delta\phi$. Then, the number of pulses interrogated is $N = T_{\mathrm{inter}}\frep = 10^5$. Using protocol 1B, this leads to an enhancement of the sensitivity by a factor $\chi_{1B}=N=10^5$, which translates in a stabilization of the frequency offset down to $\delta\nu_0 \sim  \frep/N = 1/T_{\mathrm{inter}} = 1\,$kHz.

This result can be improved by applying the protocols with two pulse sequences (2A,2B). To be specific, we can pair $N=5\times10^4$ pulses with delays $N_d=5\times10^4$ and reach, with protocol 2B, a resolution $\delta\nu_0=\frep/(N N_d) = 40\,$mHz. Let us note that duty cycles (i.e., the time required for ion Doppler cooling + probing $\Delta\phi$ + ion-state detection + ion-state reset) of $\sim1$~ms have been reported~\cite{hemmerling2012}, so assessing the precision using $1\,$ms for the interrogation time can be considered a conservative estimate.

Again, we remark that the numerical estimates in the main text of the paper and this supplementary material take into consideration only the coherence properties of trapped ions for the stabilization of an ``ideal frequency comb'', and technical issues inherent to currently available combs are not included in the calculations.

\subsubsection{Sensitivity enhancement with self-referenced Raman schemes}

Use of a Raman scheme lifts the restrictions related to the excited-state lifetime of the qubit as spontaneous decay is of no concern.  Such a scheme has been implemented with various systems, e.g., \Yb\ with a qubit defined by the $^2S_{1/2}$ hyperfine states $|F=1,m_F=0\rangle=|1\rangle$ and $|F=0,m_F=0\rangle=|0\rangle$, which are split by $\omega_{at}=12.642815$~GHz~\cite{Campbell2010,Senko2012}. For these states, coherence times larger than 1 second have been measured~\cite{Olmschenk2007}.

Let us consider a pulse train of $1\,$ms, which provides $10^5$ pulses at $\frep=100\,$MHz, and let us split this train on two lines. We seek to maximize the phase difference between them. To this end, we consider the available $10^5$ pulses into sets of $10^4$ and keep the first set, ${\cal S}_1=\{1,\ldots,10^4\}$, the set ${\cal S}_2=\{10^4+1,\ldots,2\times 10^4\}$, and the last set, ${\cal S}_3=\{9\times 10^4+1,\ldots,10^5\}$. The first set, ${\cal S}_1$, will be further split in two, so that half of the pulses are paired with those in ${\cal S}_2$ ($N_{d1}=10^4$) and the other half with ${\cal S}_3$ ($N_{d2}=9\times10^4$). Then, this optical setup yields a sensitivity enhancement of order $\chi=N|N_{d2}-N_{d1}|=8\times10^8$ or $\delta\nu_0 \sim 0.1\,$Hz. If we allow ourselves a longer interrogation time of $1\,$s, the figures would improve down to an amazing precision of $10^{-7}\,$Hz.

We note that these high sensitivies are achievable almost independently of the underlying physical system used for the qubit: taking into account the continuum spectrum of each pulse, the only requirement is the proximity of $\omega_{at}$ and $\bar{\omega}$, a feature that can be engineered, and coherence times which are experimentally available.

\subsubsection{Recursive refinement for large pulse-to-pulse phase shifts}

In our studies we have found that the sensitivity of our metrology protocols can be written in the form $\sigma^{-1} \sim \chi(N)\sqrt{M}$, where the enhancement factor $\chi(N)$ arises from a clever accumulation of the phase. In practice, for a non-stabilized frequency comb with a large $\Delta\phi$ and an excessive number of pulses, the total accumulated phase, $\chi(N)\Delta\phi,$ will wrap around the maximum measurable value, $\pi$, precluding a unique determination of $\Delta\phi$.

The appropriate way to deal with this situation is to do an iterative refinement of the phase measurement. As an example, let us consider a fiber-based frequency comb: these devices have an intrinsic width of the offset frequency of about 200~kHz. This means that, for $\frep=100\,$MHz, the phase $\phi_m$ may wrap around $\pi$ in about $N=500$ pulses. Hence, on the first iteration, it is meaningless to interrogate the laser for much longer than a few $\mu$s. This iteration allows us already to achieve a precision in $\Delta\phi$ of order $\sqrt{M}/500$ with protocol 1B. This initial value can be used to fed back to the laser setup, to lower $\Delta\phi$ and, on the next iteration, use a larger number of pulses.

Continuing with this example, a similar iterative refinement using protocol 2B and a fixed interrogation time $\sim 1\,\mu$s, would lead to an accuracy in the comb offset frequency of $\delta\nu_0 = \frep/(N N_d) = \frep/250^2 = 3\,$Hz using only one ion.

\begin{figure*}[tb]
  \centering
  \includegraphics[angle=-90,width=0.85\linewidth]{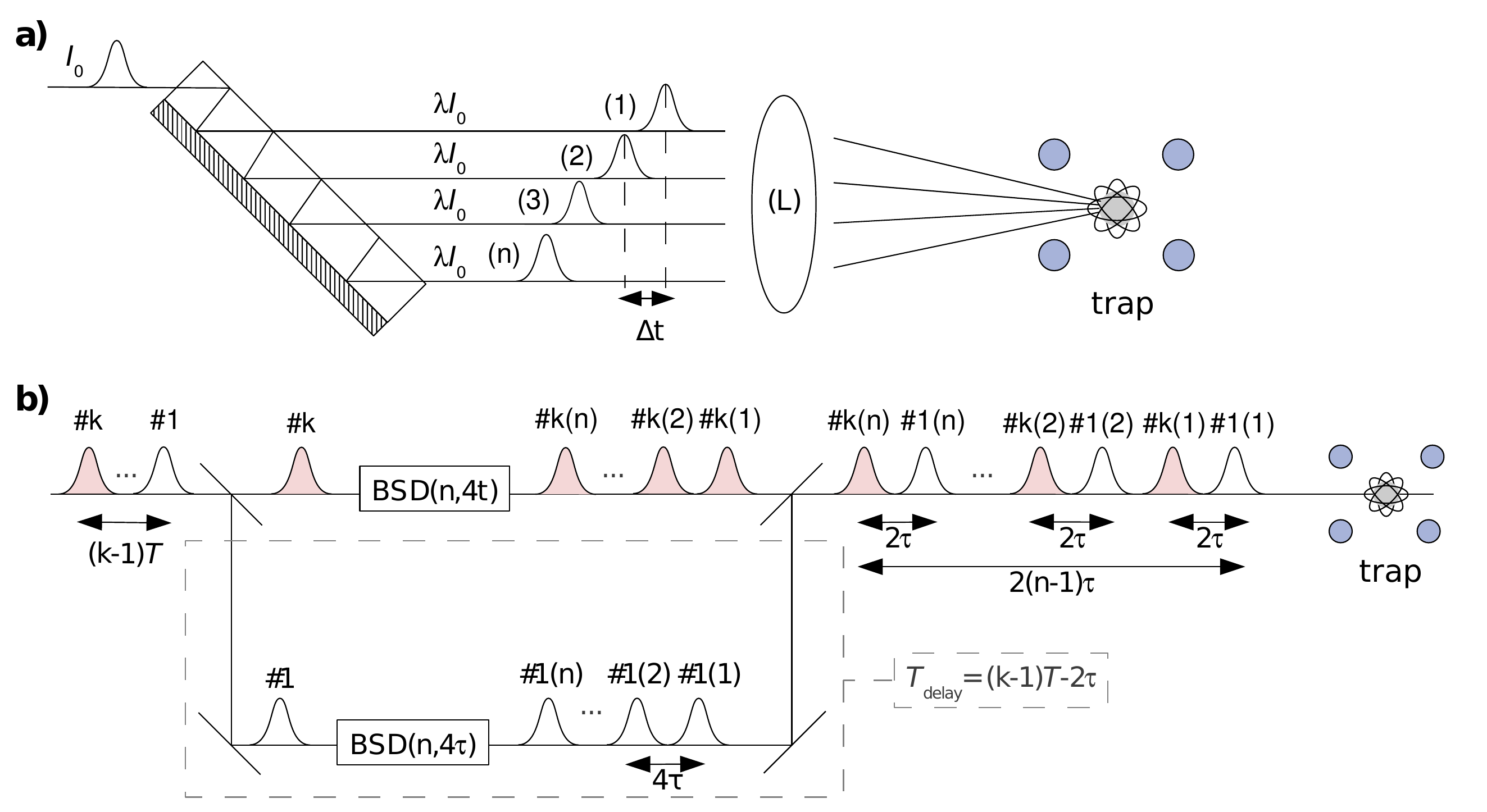}
  \caption{(a) A beam splitter and delayer (BSD) can be implemented by shining the laser pulse train into a semitransparent mirror facing a perfect mirror (shaded area). The incident angle of the beam determines the number, $n$, and delay, $\Delta t$, between the outgoing replica pulses, which are focused by a lens (L) onto the ion.
(b) Optical setup with a delay line (dashed box) and two BSDs with delay $\Delta t=4\tau$. This transforms the incident pulse train into a new train with interleaved pulses with a fixed phase difference between them and a total duration $2(n-1)\tau \ll kT$.}
  \label{fig:ultimate}
\end{figure*}

\section*{{\S}G. Ultimate precision limits with advanced pulse-shaping techniques}

The protocols discussed so far achieve a great efficiency thanks to the number of pulses in a given interrogation time and possible delays among them. Note however, that the comb is mostly ``empty'': between every two pulses of about 10~ps, there is a waiting time $\sim 10\,$ns in which the ion is idle. It would seem that this empty time, combined with the coherence rates of the ions, set the ultimate limits for precision in our setup. However, if the laser has enough power, we can engineer a clever scheme to fill these empty gaps, increasing the effective repetition rate of the ion-laser interaction.

The trick here will be to ``compress'' the pulses so that a minimal time elapses between the end of one pulse and the beginning of the following one, but without modifying the phase of any one pulse.  Such ``compression'' could be realized with an optical setup as depicted in Fig.~\ref{fig:ultimate}b. The key ingredient in this setup is an optical device which we call Beam Splitter and Delayer (BSD) that, given an intense ultrashort pulse, extracts a train of $n$ replica pulses separated by a very short time $\Delta t$ [cf.~Fig.~\ref{fig:ultimate}a]. (An alternative BSD optical setup producing 8 replicas of an initial pulse has been recently implemented in~\cite{Senko2012}.)

We will discuss these ideas in a particular application: doing metrology of the comb with a dipole transition. In this case the qubit of choice will not satisfy the condition $\taucoh \gg T$. Consider for example Ca$^+$ ions using the dipole-coupled $|S_{1/2}\rangle$ and $|P_{1/2}\rangle$ states for which $\taurad \approx 7$~ns~\cite{Arora2007}. In this setup we can still reach high precisions taking a relatively long pulse train of duration $\gg \taucoh$ as long as we ensure that all the pulses pass through the ion within a short time $\lesssim \taucoh$.  On the other hand, we must still fulfill the requirement that different pulses do not overlap in time, that is $\Delta t \ge 2\tau$, which restricts us to use sets of up to $n \le \mathrm{min}\{\taurad,\taucoh\}/(2\tau) \sim 7~\mathrm{ns}/20~\mathrm{ps}=350$ replica pulses. To be concrete, let us use the setup in Fig.~\ref{fig:ultimate}b to pick up two pulses with a relative delay of $T_d=N_d T = 10\,\mu$s ---this corresponds to pulses $1$ and $k=N_d=10^3$ in the previous figure. The pulses will go through the BSD and be recombined, alternating replica pulses from each line. For a conservative $n=4$ (not to lose too much power in each replica), $\Delta t=4\tau$, and $N_d \approx 1000$, we obtain a phase sensitivity enhancement by a factor $n N_d \approx 4\times 10^3$.

The same ideas can be applied to the Raman scheme by ensuring that the replica pulses from both lines arrive simultaneously to the ion. The result is an enhancement of the sensitivity by an additional factor $n$ on top of the formulae derived in the main text of the paper.

We finally note that the very short probe times considered here, allow for the recollection of a large set of statistical data in a very short time. Together with the large sensitivity enhancements calculated, the presented schemes appear as very competitive protocols to measure and stabilize the carrier-envelope offset phase of frequency combs without the need for octave-spanning spectra.

\section{\label{sec:candidates}Frequency comb technologies and candidate systems for their stabilization}

To conclude, let us discuss in some more detail a couple of contexts where present frequency comb technology might benefit particularly from the protocol we propose.
\begin{table}
\caption{Regions of the electromagnetic spectrum where the phase stabilization protocol may be immediately applicable with the systems discussed in {\S}H.}
\label{tab:spectrum}
\begin{tabular}{lrrl}
  Region  & Wavelength & Energy  & Interest \\
  \hline
  Near-UV  & 10-400  nm & 3-124 eV           & HHG, strong fields\\
  Visible     & 390-750 nm & 430-790 THz    & Electronic transitions \\
  Near-IR   & 0.8-2.5 $\mu$m & 120-430 THz    & Vibrational overtones \\
  Mid-IR     & 2.5-25 $\mu$m & 12-120 THz  & Rot-vib structure
\end{tabular}
\end{table}
The stabilization of frequency combs (FCs) in the optical frequency range (cf. Table~\ref{tab:spectrum}) is nowadays quite well solved with several protocols, and there are also various technologies that enable their optical and spectral manipulations.
The situation is not so advanced in other regions of the spectrum which nevertheless are of high relevance for several scientific and technological applications.
For example, the mid-infrared (mid-IR) frequency range is where many characteristic molecular vibrational and rotational lines lie, which makes it of biological, chemical and physical interest for molecular detection and trace analysis. In addition, the atmosphere is relatively transparent at these wavelengths, which makes them valuable for astronomical studies.
For these reasons, in the last years there is a growing interest in developing FCs in this spectral region~\cite{schliesser2012}. Several technologies are being developed to realize these FCs, such as mode-locked lasers, difference-frequency generation (DFG), optical parametric oscillators (OPOs), and microresonator-based Kerr combs~\cite{schliesser2012}.
Let us focus on DFG.

\begin{figure}[tb]
  \centering
  \includegraphics[width=0.45\linewidth]{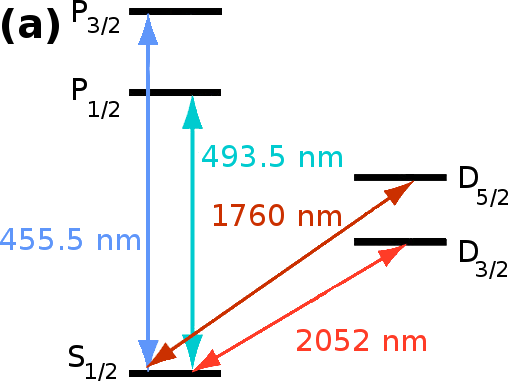}\hfill%
  \includegraphics[width=0.4\linewidth]{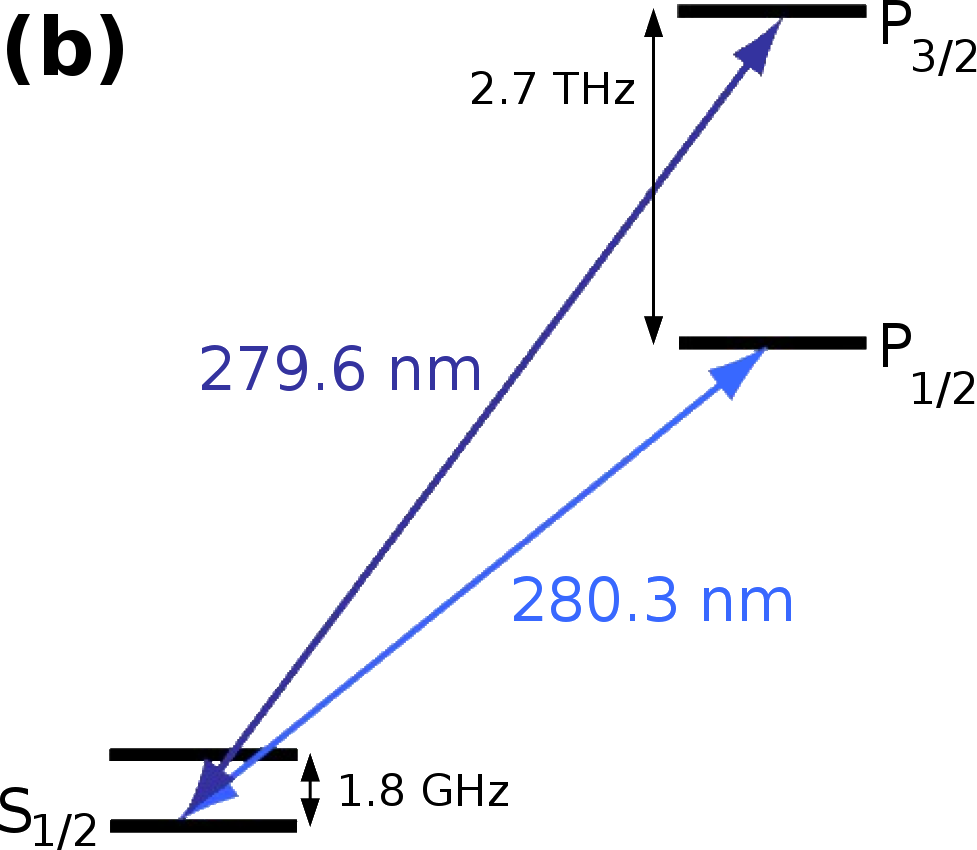}
  \caption{Lowest energy levels of (a) $^{137}$Ba$^+$ and (b) $^{25}$Mg$^+$.}
  \label{fig:candidates}
\end{figure}

Here, one uses a nonlinear optical effect to transfer energy from the visible or near-IR into the mid-IR. For example, one can take a near-IR FC with frequencies $\nu_n = n\frep + \fceo$ and mix it with a CW laser of frequency $\fcw$. Then, a new comb with frequencies $\nu^{\mathrm{DFG}}_{n} = |\nu_n - \fceo |$ is obtained.
Achieving phase matching on all the desired bandwidth can be eased
by using either two stabilized FCs, or two teeth of a single comb.
One gets in the latter case $\nu^{\mathrm{DFG}}_{n,m}=|n-m|\frep$.
It is clear that this approach can benefit from the protocol that we propose if one has access to a probe ion sensitive to mid-IR frequencies close to the desired range of $\nu^{\mathrm{DFG}}_{n,m}$. A good candidate for this can be Ba$^+$, whose lower electronic states we show in Fig.~\ref{fig:candidates}(a). We see that the transitions from the ground electronic state, S$_{1/2}$, to the long-lived, metastable D$_{3/2,5/2}$ states have wavelengths in the near-IR ($\lambda=1760$, 2052~nm). 
These states are very long-lived (with lifetimes $\tau \gtrsim 80$~s and $\approx30$~s, respectively~\cite{safro2010-Ba}), and one could use them to encode the qubit \ket{1} state, setting \ket{0}=S$_{1/2}$, in a similar way as is done with Ca$^+$. Experiments with this ion~\cite{myerson2008det} reached very high detection fidelities ($99.99\%$) with detection times of $145~\mu$s using an adaptive measurement technique via the S$_{1/2}\to P_{1/2,3/2}$ transitions ($\tau_{1/2, 3/2}=7.9,~6.3$~ns, resp.); recent improvements on this method with Mg$^+$~\cite{hemmerling2011,hemmerling2012} and Yb$^+$~\cite{noek2013} have reached high detection efficiencies $\geq 95\%$ (sufficient for our needs) in times as short as $10~\mu$s. We note that the typical phase noise of unstabilized optical FCs has a growing spectral weight in the region $\sim 0.1-1$~kHz~\cite{Steinmeyer2005}, which can be stabilized with interrogation times $\simeq 10$~ms, much shorter than the ion state-detection timescale.

On the other side of the visible frequencies, there is also a growing number of laboratories working in the UV region of the spectrum, with research on strong-field physics, high-harmonic generation (HHG), ultra-fast processes, above-threshold ionization, and others.
The well-known $f$-$2f$ technique has problems in the UV due to the difficulty of frequency-doubling at such high energies, for example because of the damage of the nonlinear material.
Here, a suitable ion to implement the phase-stabilization protocol could be magnesium, that features two lines around 280~nm, see Fig.~\ref{fig:candidates}(b). In this case, the excited states have lifetimes of 3.8~ns~\cite{ansbacher1989,safro1998-Mg,hemmerling2011}. This limits the duration of the direct interrogations, and points rather to the Raman schemes, which can be implemented using two different hyperfine states of the ground electronic state as the qubit states, with the P$_{1/2,3/2}$ states playing the role of the off-resonant excited state \ket{e} discussed in the main text.

%
\end{document}